\documentclass[prb,twocolumn]{revtex4-1}

\usepackage{graphicx}
\usepackage{amsmath,amssymb,bm}
\usepackage{dcolumn}
\usepackage{booktabs}
\usepackage{subfig}
\usepackage{xcolor}
\usepackage{array}

\usepackage{tikz}
\usetikzlibrary{arrows.meta,decorations.pathmorphing}

\newcommand{\ket}[1]{|#1\rangle}

\newcolumntype{M}[1]{>{\centering\arraybackslash}m{#1}}

\renewcommand{\Im}{\mathrm{Im}\,}

\newcommand{\p}{\partial}

\def\be{\begin{equation}}
\def\ee{\end{equation}}
\def\bearr{\begin{eqnarray}}
\def\eearr{\end{eqnarray}}

\usepackage[colorlinks=true,linkcolor=red]{hyperref}

\begin{document}
	
	\title{A Quantum Framework for Negative Magnetoresistance in Multi-Weyl Semimetals}
	\bigskip
	\author{Arka Ghosh, Sushmita Saha and Alestin Mawrie}
	\normalsize
	\affiliation{Department of Physics, Indian Institute of Technology Indore, Simrol, Indore-452020, India}
	\date{\today}
\begin{abstract}
We develop a fully quantum-mechanical theory of negative magnetoresistance in multi-Weyl semimetals in the ${\bf E}\parallel{\bf B}$ configuration, where the chiral anomaly is activated. The magnetotransport response is governed by Landau quantization and the emergence of multiple chiral Landau levels associated with higher-order Weyl nodes. These anomaly-active modes have unidirectional dispersion fixed by the node’s monopole charge and dominate charge transport. As the magnetic field increases, individual chiral branches successively cross the Fermi energy, producing discrete slope changes in the longitudinal conductivity and a step-like negative magnetoresistance. This quantized evolution provides a direct experimental signature of multi-Weyl topology. Bulk Landau levels contribute only at very low fields due to strong disorder scattering and do not affect the anomaly-driven regime. Our results establish a unified, fully quantum-mechanical framework in which negative magnetoresistance arises from the discrete Landau-quantized spectrum and microscopic impurity scattering, beyond semiclassical anomaly descriptions.
\end{abstract}

\email{amawrie@iiti.ac.in}
	
\maketitle

\section{Introduction}

Negative magnetoresistivity (NMR) in the ${\bf E}\!\parallel{\bf B}$ geometry is a
well-established transport signature of the chiral anomaly in Weyl and 
multi-Weyl semimetals.  
When electric and magnetic fields are aligned, the anomaly (the 
nonconservation of chiral charge induced by parallel ${\bf E}$ and ${\bf B}$) pumps 
charge between Weyl nodes of opposite chirality, thereby enhancing the 
longitudinal conductivity and producing a characteristic decrease in 
resistivity~\cite{son2013chiral,burkov2015qm,zrj2015nmr,kim2013dirac,huang2015nmr}.  
Despite its central role, most theoretical descriptions of NMR remain 
classical or semiclassical in nature, relying on Berry-curvature-modified 
Boltzmann equations, hydrodynamic anomaly relations, or semiclassical kinetic 
theory~\cite{son2013chiral,spivak2016hydro}.  
Such approaches capture the anomaly-driven drift but do not resolve the 
microscopic Landau-level structure or the detailed influence of impurity 
scattering.  
A fully quantum framework is therefore required to understand how chiral 
Landau levels evolve with magnetic field, how their degeneracies are modified, 
and how their scattering rates shape the NMR response.

Weyl semimetals host a pair of band-touching points with linear or nonlinear 
dispersion~\cite{wan2011wsmdirac,burkov2011weylmetal}.  
Each Weyl node acts as a quantized monopole of Berry curvature and carries a 
topological charge that governs anomalous transport and magneto-optical 
phenomena~\cite{armitage2018review,yan2017topological}.  
Under a magnetic field this topological structure enforces the existence of a 
chiral Landau level with strictly unidirectional dispersion, which underlies 
the anomaly-induced enhancement of longitudinal conductivity in the 
${\bf E}\!\parallel{\bf B}$ configuration~\cite{nielsen1983adlerbell, burkov2014ahe}.

Multi-Weyl semimetals (mWSMs) provide a natural generalization in which the 
nodes carry monopole charge $m>1$, protected by crystalline rotational 
symmetries~\cite{fang2012multiweyl,fang2012multiweyl_1}.  
Their transverse dispersion becomes nonlinear, and magnetic quantization yields 
$m$ chiral Landau levels~\cite{Y_Sun, fang2012multiweyl}.  
The presence of multiple anomaly-active channels strengthens and reshapes the 
longitudinal transport response, thereby should give rise to enhanced, and in many cases 
multi-step-NMR profiles. 
Candidate platforms include SrSi$_2$, EuIn$_2$As$_2$, strained MoTe$_2$ and 
WTe$_2$, the LaAlGe family, kagome magnets such as Co$_3$Sn$_2$S$_2$, and 
chiral crystals such as AlPt and 
AlPd~\cite{H_Weng_PRX,Y_Xu_prl,S_Kimura_prb,xu2017laalge,chang2016laalge,liu2018co3sn2s2,wang2018co3sn2s2,chang2018alpt,schroter2019topological, schroter2019topological,Lv2015TaAs, NXu2015TaAs, Yang2015NbP, Huang2016SrSi2, Tang2017CoSi, Schroter2019RhSi}.  
In these materials, the spacing between successive chiral Landau levels,
$\Delta E_{\mathrm{ch}} = E_{n+1}^{\mathrm{ch}} - E_{n}^{\mathrm{ch}}$,
is set by magnetic-field-controlled Landau quantization and remains comparable
to or larger than the thermal broadening $k_{\mathrm{B}}T$ over the
experimentally relevant field range $0.5 < B < 4.5~\mathrm{T}$, as established
for realistic Weyl and tilted Weyl systems in Refs.~[\cite{tchoumakov2016tilt, soluyanov2015typeII}].  
Thus, even at room temperature the chiral levels are well resolved, and 
transport along ${\bf E}\!\parallel{\bf B}$ proceeds through discrete quantum 
channels rather than semiclassical or hydrodynamic drift.  
This hierarchy provides strong motivation for a unified, fully quantum description of NMR across different topological charges.

While semiclassical formulations such as those of 
Son \& Spivak~\cite{son2013chiral} and 
Spivak \& Andreev~\cite{spivak2016hydro} 
successfully predict a linear-in-$B$ magnetoconductivity, they treat the 
anomaly-induced current as a continuum drift process and therefore cannot 
capture the microscopic structure that emerges under Landau quantization.  
In particular, semiclassical theories do not resolve  
(\textbf{\textit{i}}) the discrete set of $m$ chiral Landau levels fixed by the monopole charge,  
(\textit{\textbf{ii}}) the field-dependent depopulation of individual chiral branches, or  
(\textbf{\textit{iii}}) the resulting sequence of kinks and piecewise-linear segments in 
$\sigma_{zz}(B)$.  
The present work goes beyond these approaches by developing a fully quantum, 
Landau-level-resolved theory of magnetotransport in which the $m$ chiral modes 
govern the anomaly-induced conductivity.  
A central result is the identification of characteristic magnetic fields $B_{n}$ 
at which the $n^{\rm th}$ chiral Landau level crosses the Fermi energy and ceases to 
contribute to transport.  
This mechanism produces the multi-step structure of $\sigma_{zz}(B)$ and the 
corresponding kinks in the magnetoresistance-direct fingerprints of the node’s 
topological charge with no analogue in semiclassical theories.  
Although both semiclassical and quantum approaches predict an overall negative 
magnetoresistance, the microscopic origin, internal structure, and field-dependent 
signatures identified here are uniquely quantum and arise directly from the 
Landau-quantized spectrum of multi-Weyl fermions.

In this work, we develop such a fully quantum-mechanical framework for NMR in 
Weyl and multi-Weyl semimetals.  
Our theory incorporates the complete Landau-level spectrum, screened Coulomb 
disorder, and the Kubo formalism, enabling a transparent treatment of both 
chiral and bulk channels.  
We show that the anomaly-induced NMR profile is dictated by the number of 
chiral branches: triple-Weyl semimetals ($m=3$) exhibit three distinct linear 
regimes in $\sigma_{zz}(B)$, double-Weyl systems ($m=2$) display two, and 
ordinary Weyl nodes ($m=1$) show a single smooth linear increase without 
intermediate structure.  
Whenever a chiral branch is pushed above the Fermi level, its contribution 
vanishes abruptly, producing an observable change in slope.  
These features emerge naturally within our quantum treatment and cannot be 
reproduced using semiclassical anomaly theories.

The remainder of this paper presents the full derivations, demonstrating how 
topology, Landau quantization, and microscopic impurity scattering combine to 
produce the chiral-anomaly-induced NMR response in Weyl and multi-Weyl 
semimetals.

\section{Hamiltonian Formulation and Emergence of Chiral States}

The low-energy excitations of Weyl and multi-Weyl semimetals are governed by 
topologically protected band-touching points whose monopole charge determines the 
degree of anisotropy and non-linearity in their dispersion.  
For a multi-Weyl node of order $m$, the minimal continuum Hamiltonian can be written as
\begin{equation}
H_0=\frac{\lambda}{2}\left(k_-^{m}\sigma_+ + k_+^{m}\sigma_- \right)
+ \eta v_z\left(k_z + \eta\frac{\mathcal{Q}}{2}\right)\sigma_z ,
\label{eq:H0}
\end{equation}
where $\sigma_{x,y,z}$ are the Pauli matrices acting on the pseudospin basis,
$\sigma_\pm=\sigma_x\pm i\sigma_y$, and $k_\pm=k_x\pm i k_y$.  
The parameter $\lambda$ determines the strength of the in-plane nonlinearity, 
$v_z$ is the band velocity along $k_z$, and $\mathcal{Q}$ is the momentum-space separation between the two Weyl nodes of chirality $\eta=\pm1$.  The integer $m$ plays the role of the monopole charge of the Berry curvature and determines both the topological and spectroscopic properties of the node.\cite{fang2012multiweyl}
Also, this Hamiltonian correctly reproduces the anisotropic dispersion observed in density-functional calculations and tight-binding models of multi-Weyl systems\cite{fang2012multiweyl,s_saha}.

\medskip\noindent\textit{\textbf{Inclusion of a tilt and quadratic particle-hole asymmetry term}}:
In many Weyl and multi-Weyl materials, the dispersion around the node is not perfectly upright but acquires a finite tilt along a crystallographic direction. This tilt preserves the Weyl point but reshapes the energy cone and, when sufficiently strong, drives a Type-I to Type-II transition marked by touching electron and hole pockets.\cite{soluyanov2015typeII,tchoumakov2016tilt}
Given its pronounced impact on the density of states and transport coefficients, the tilt must be included for a realistic description. In addition to the linear Weyl dispersion, realistic band structures generally exhibit particle-hole asymmetry arising from higher-order momentum terms. This effect is modeled by including a quadratic contribution in the transverse plane, which preserves the Weyl point and does not affect its topology, but modifies the Landau-level spacing and transport response. Such particle-hole asymmetric corrections are commonly retained in low-energy descriptions of tilted and non-ideal Weyl systems.\cite{tchoumakov2016tilt, soluyanov2015typeII} Below we incorporate the above correction terms into the Hamiltonian and study how it modifies both the chiral Landau levels and the global Landau-level structure.
\begin{equation}
H_{\rm tilt}
=w_{\parallel}(k_x^2+k_y^2)\sigma_0 + w_z k_z\sigma_0 ,
\end{equation}
which does not gap out the Weyl point but shifts and skews the cones.  Here 
\(H_{\rm tilt}\), collectively denotes the scalar terms responsible for the longitudinal tilt of the Weyl cones and the particle-hole asymmetry of the spectrum.
We tabulate the system parameters as follows [\ref{tab:params}]
\begin{table}[h!]
\renewcommand{\arraystretch}{1.3}

\begin{center}
\begin{tabular}{c|c c c c}
\hline\hline
\hspace{0mm}\textbf{Monopole charge}\hspace{4mm} &
\hspace{1mm}\textbf{$v_{z}$}\hspace{4mm} &
\hspace{1mm}\textbf{$w_{z}$}\hspace{4mm} &
\hspace{1mm}\textbf{$\lambda$}\hspace{4mm} &
\hspace{1mm}\textbf{$w_{\parallel}$}\hspace{4mm}
\\
\hline
$\mathbf{m=1}$ & $1.0$ & $0.25\,v_{z}$ & $0.30$ & $0.20$ \\
$\mathbf{m=2}$ & $1.5$ & $0.25\,v_{z}$ & $0.30$ & $0.8\,\lambda$ \\
$\mathbf{m=3}$ & $2.0$ & $0.25\,v_{z}$ & $0.60$ & $0.30$ \\
\hline\hline
\end{tabular}
\end{center}
\caption{ Model parameters used in this work. 
The chosen values lie within experimentally measured and first-principles-calculated ranges for Weyl and multi-Weyl semimetals. For $m=1$, typical band velocities and tilt strengths are guided by ARPES and \emph{ab initio} studies of TaAs and NbP (see Refs.[~\cite{Lv2015TaAs,NXu2015TaAs,Yang2015NbP}]).  For $m=2$, the nonlinear dispersion parameter $\lambda$ and the anisotropy $v_{z}/\lambda$ are taken to be consistent with double-Weyl candidates such as SrSi$_2$ from first-principles calculations~\cite{Huang2016SrSi2}.  
For $m=3$, the enhanced anisotropy and nonlinear coefficients reflect the cubic 
Weyl dispersion observed in multifold chiral crystals such as CoSi and RhSi, whose velocities and node separations have been established by ARPES and DFT (Refs.[\cite{Tang2017CoSi,Schroter2019RhSi}]). These references show that the adopted dimensionless ratios $w_{z}/v_{z}$, $\lambda$, and $w_{\parallel}$ fall within realistic material-specific ranges, making the present model quantitatively representative of experimentally relevant multi-Weyl systems.}
\label{tab:params}
\end{table}

\medskip\noindent\textit{\textbf{Landau quantization}}:  
To expose the chiral structure of the spectrum, we apply a magnetic field 
$\mathbf{B}=B\hat{z}$ and adopt the Landau gauge 
$\mathbf{A}=(-By,0,0)$.  
The canonical momenta transform as
\begin{equation}
\Pi_x = p_x - By,\qquad \Pi_y = p_y ,
\end{equation}
and it is convenient to introduce the standard ladder operators
\begin{equation}
a=\frac{\Pi_x - i\Pi_y}{\sqrt{2B}},\qquad
a^\dagger=\frac{\Pi_x + i\Pi_y}{\sqrt{2B}},
\end{equation}
which satisfy $[a,a^\dagger]=1$.  
In terms of these operators, the total Hamiltonian in Eq.~(\ref{eq:H0}) becomes\cite{soluyanov2015typeII,s_saha}
\begin{eqnarray}
&&H_T=2Bw_\parallel\left(a^\dagger a+\frac{1}{2}\right)\sigma_0\nonumber\\&&+
\begin{bmatrix}
(w_z+\eta v_z)(k_z+\eta \frac{\mathcal{Q}}{2}) &
\lambda_m a^m \\[4pt]
\lambda_m (a^\dagger)^m &
(w_z-\eta v_z)(k_z+\eta \frac{\mathcal{Q}}{2})
\end{bmatrix},
\end{eqnarray}
with $\lambda_m=\lambda(2B)^{m/2}$.
This form highlights the fundamental distinction between single ($m=1$) and multi-Weyl ($m>1$) nodes:  
the magnetic field couples Landau oscillator states $|n\rangle$ to $|n-m\rangle$, 
rather than to $|n-1\rangle$, reflecting the $m$-fold vortex structure in the transverse dispersion. 
For $n\ge m$, we thus have the trial spinor
\begin{equation}
\psi_{n,k_z}^{\eta,s}=
\begin{pmatrix}
\alpha_n^{\eta,s}(k_z) |n-m\rangle\\[4pt]
\beta_n^{\eta,s}(k_z) |n\rangle
\end{pmatrix},
\label{eq:spinor_bulk}
\end{equation}
which yields the bulk Landau-level energies
\begin{equation}
E_n^{\eta,s}(k_z)
=(2n-m+1)Bw_\parallel
+\hbar w_z\!\left(k_z+\eta\frac{\mathcal{Q}}{2}\right)
+s\,\Gamma_n^m,
\label{eq:bulk_dispersion}
\end{equation}
with
\begin{equation*}
\Gamma_n^m
=
\sqrt{
\left[
-mBw_\parallel
+\eta v_z\!\left(k_z+\eta\frac{\mathcal{Q}}{2}\right)
\right]^2
+\frac{n!}{(n-m)!}\lambda_m^2
}.
\end{equation*}
 Also,
\begin{equation}
\left.
    \begin{aligned}
        \alpha_{n}^{\eta,s}(k_z)=\frac{1}{\sqrt{2}}\sqrt{1-s\frac{Bmw_\parallel-\eta v_z(k_z+\eta \frac{\mathcal{Q}}{2})}{\Gamma_n^m}}\\
        \beta_{n}^{\eta,s}(k_z)=\frac{1}{\sqrt{2}}\sqrt{1+s\frac{Bmw_\parallel-\eta v_z(k_z+\eta \frac{\mathcal{Q}}{2})}{\Gamma_n^m}}
    \end{aligned}
    \right\}.
\end{equation}
These modes are gapped away from the chiral Landau states and possess wavefunctions containing higher-order Hermite polynomials, whose increased oscillatory structure enhances their overlap with short-range disorder, making them more susceptible to impurity scattering.

\medskip\noindent\textit{\textbf{Emergence of chiral states}}:  
For $n<m$, one cannot form the state $|n-m\rangle$, and the upper spinor component in 
Eq.~(\ref{eq:spinor_bulk}) vanishes identically.  
The eigenvalue equation reduces to a single-component problem, producing
\begin{equation}
E_n^{\eta,\rm ch}(k_z)
=(w_z-\eta v_z)\!\left(k_z+\eta\frac{\mathcal{Q}}{2}\right)
+(2n+1)Bw_\parallel.
\label{eq:tilted_chiral}
\end{equation}
with eigen-spinor equal to $(0,|n\rangle)^{\mathrm{T}}$.  
These $m$ modes constitute the celebrated chiral Landau levels, whose propagation direction is locked to the chirality $\eta$ of the Weyl node\cite{Yuan2018,Zhao2021}.  
This unique combination of features underlies the Adler-Bell-Jackiw chiral anomaly \cite{zrj2015nmr, kim2013dirac, nielsen1983adlerbell}
and is responsible for the characteristic linear in $B$ enhancement of longitudinal conductivity in the longitudinal ($\mathbf{E}\parallel\mathbf{B}$) configuration.
\begin{figure}[t]
    \includegraphics[width=89mm,height=70.5mm]{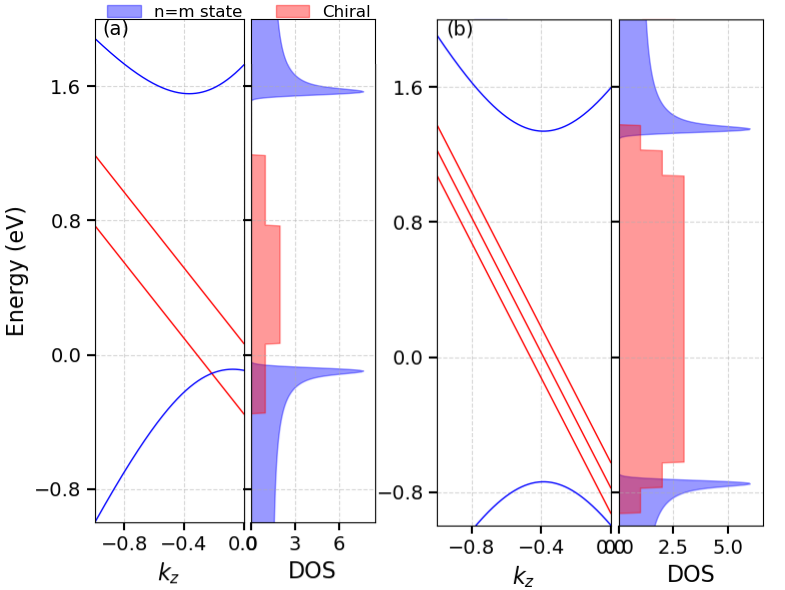}     \caption{ Energy dispersion and density of states (DOS) of Landau-quantized multi-Weyl semimetals in a magnetic field. Panel (a) shows a double-Weyl node ($m=2$) and panel (b) a triple-Weyl node ($m=3$). In each case, the left plot displays the Landau-level spectrum versus $k_z$, while the right plot shows the corresponding DOS. Red lines denote the $m$ chiral Landau levels with unidirectional dispersion fixed by the node chirality, and blue curves represent bulk ($n=m$) Landau levels.}

\label{fig:DOS_m2_m3}
\end{figure}
These chiral modes remain unidirectional, robust, and topologically protected as long as the tilt does not exceed the critical threshold separating type-I and type-II Weyl phases\cite{soluyanov2015typeII,tchoumakov2016tilt}.  
Their presence and evolution with magnetic field are central to the quantum description of longitudinal magnetotransport and the emergence of NMR analyzed in later sections.

\medskip\noindent\textit{\textbf{Density of States of Chiral States}}:  
Mathematically, the DOS of chiral states for Weyl semimetal with topological charge \(m\) can be written as,
\begin{eqnarray}
   \mathcal{D}^\eta_{\rm ch}(E)&&=\frac{1}{2\pi {l_B}^2} \sum_{n=0}^{m-1}\int\frac{dk_z}{2\pi}\delta(E-E_n^{\eta,\rm ch}(k_z))\nonumber\\&&\times\Theta(E-E_n^{\eta,\rm min})\Theta(E_n^{\eta,\rm max}-E),
\end{eqnarray}
where the magnetic length \(l_B=1/\sqrt{B}\).
Also, we defined $E_n^{\eta,\rm min}=E_n^{\eta,\rm ch}(k_z=0)$ and $E_n^{\eta,\rm max}=E_n^{\eta,\rm ch}(k_z=\mathcal{Q})$ to be the maximum and minimum energy value of an \(n^{\rm th}\) chiral state, respectively, defined as 
\begin{equation}
\left.
\begin{aligned}
    E_n^{\eta,\rm min} &= (2n+1)w_\parallel  B - (v_z - \eta w_z)\frac{\mathcal{Q}}{2},\\[4pt]
    E_n^{\eta,\rm max} &= (2n+1)w_\parallel B
    + \mathcal{Q}\left(1+\frac{\eta}{2}\right)\left(w_z-\eta v_z\right)
\end{aligned}
\right\},
\end{equation}
which specify the lowest and highest possible energy along the \(n^{\rm th}\) chiral 
channel. These bounds arise because the chiral Landau level disperses 
linearly in $k_{z}$ with slope determined by the combination $(w_{z}-\eta v_{z})$, and terminates at the locations of the two Weyl nodes separated by momentum $\mathcal{Q}$. We derived the above quantities by adopting a fully linear dispersion (Eq. [\ref{eq:tilted_chiral}]) over the interval $0<k_{z}<\mathcal{Q}$ because the experimentally relevant transport window lies entirely within the linear regime of the Weyl cone, well before lattice-scale curvature becomes appreciable near the two Weyl nodes at \(k_z=0\) and \(k_z=\mathcal{Q}\).

After integration over the $k_z$ space, we get the DOS of chiral state to be,
\begin{eqnarray}\label{Dos_Chi}
 &&   \mathcal{D}^\eta_{\rm ch}(E)=\frac{1}{4\pi^2 l_B^2|w_z-\eta v_z|}\nonumber\\&\times &\sum_{n=0}^{m-1}\Theta(E-E_n^{\eta,\rm min})\Theta(E_n^{\eta,\rm max}-E).
\end{eqnarray}

\noindent This result (in Fig. [\ref{fig:DOS_m2_m3}]) shows that the DOS of the chiral states is constant within each allowed energy window and inversely proportional to the chiral velocity $|w_{z}-\eta v_{z}|$. The existence of $m$ distinct windows directly reflects the topological charge of the multi-Weyl node: higher monopole charge produces multiple chiral branches, each contributing a finite DOS. The presence of a nonzero chiral DOS, even at energies where the bulk DOS vanishes, is a key signature of Weyl physics in a magnetic field and plays an essential role in magnetotransport and magneto-optical phenomena such as the chiral anomaly and field-induced absorption features.

\section{Negative Magnetoresitance}

The interplay between Landau quantization and impurity scattering plays a central role in 
determining the transport properties of Weyl and multi-Weyl semimetals. In particular, when a 
magnetic field is applied parallel to the current direction, the emergence of chiral Landau 
levels leads to unconventional charge transport and a characteristic NMR. Understanding this behavior requires a detailed evaluation of the 
longitudinal conductivity, in which both the structure of the Landau-level wavefunctions and 
the corresponding scattering mechanisms-computed earlier through the transport relaxation time 
$\tau^{\mathrm{tr}}$ (See Appendix[\ref{app:selfenergy}]) enter on equal footing. 
In this section, we combine the Kubo formalism to  obtain analytical expressions for the conductivity originating from both chiral and bulk Landau levels, and we analyze how their distinct wavefunction profiles control the overall magnetotransport response.

\noindent \textbf{\textit{{Longitudinal Conductivity}}}:
To understand charge transport in a magnetic field, we compute the longitudinal conductivity 
$\sigma_{zz}$ using linear-response theory. In a Landau-quantized system, the current response 
to an applied electric field is determined not only by the band dispersion but also by the 
wavefunction structure of the Landau levels. The Kubo formula provides a convenient and fully 
quantum-mechanical framework to incorporate these effects, including disorder through the 
Green’s functions.

We begin with the standard Kubo expression for dc conductivity 
\cite{flensberg,rammer,kubo_greenwood}:
\begin{equation}\label{generic_s_zz}
    \sigma_{zz}
    =\frac{2\pi e^2 \hbar}{V}
    \int_{-\infty}^{\infty} dE
    \left[ -\frac{\partial f(E)}{\partial E} \right]
    \mathrm{Tr}\!\left[ v_z G^R(E)\, v_z G^A(E) \right],
 \end{equation}
where $G^{R/A}$ denote the retarded and advanced Green's functions, $f(E)$ is the Fermi-Dirac 
distribution, and the velocity operator along the field direction is
\(    v_z={\partial H}/{\partial k_z}\).

In the Landau-level (LL) basis $|n,k_z,s\rangle$, the Green function takes the 
diagonal form  
\begin{equation}
    G^{R/A}(E)
    =
    \sum_{n,k_z,s}
    \frac{|n,k_z,s\rangle\!\langle n,k_z,s|}
    {E-E_{n}^s(k_z)\pm i\Gamma_n^s(k_z)} ,
\end{equation}
where $\Gamma_n^s(k_z)=1/[2\tau_n^s(k_z)]$ represents the LL broadening and 
$\tau_n$ is the total scattering time associated with level $n$.  
Using this representation, the trace in Eq.~\eqref{generic_s_zz} becomes  
\begin{equation}
    \mathrm{Tr}\!\left[
        v_z G^R v_z G^A
    \right]
    =
    \sum_{n,k_y,k_z,s}
    |v_{z;n,s}(k_z)|^2\,
    G^R_{n,s}(E;k_z)\,
    G^A_{n,s}(E;k_z),
\end{equation}
where \(|v_{z;n,s}(k_z)|^2=|\langle n,k_z,s|v_z|n,k_z,s\rangle|^2\). 
The product of Green functions evaluates to  
\begin{equation}
    G^R_{n,s}(E;k_z)\,G^A_{n,s}(E;k_z)
    =
    \frac{1}
    {(E-E_{n}^s)^2+(\Gamma_n^s)^2}.
\end{equation}
We assume the width, \(\Gamma_n\) is small compared to the inter-LL spacing, the energy integral in  
Eq.~\eqref{generic_s_zz} may be performed using the Lorentzian identity  
\begin{equation}
    \int_{-\infty}^{\infty}
    dE\, 
    \frac{-\partial f/\partial E}
    {(E-\varepsilon)^2+\Gamma^2}
    \simeq
    \frac{\pi}{\Gamma}
    \left[
        -\frac{\partial f}{\partial E}
    \right]_{E=\varepsilon}
    =
    2\pi\tau_n^s
    \left[
        -\frac{\partial f}{\partial E}
    \right]_{E=\varepsilon},
\end{equation}
thereby producing a factor proportional to the LL lifetime $\tau_n^s$.  
After performing the trace over the LL guiding-center quantum number $k_y$, 
which carries the degeneracy $L_xL_y/(2\pi\ell_B^2)$, and converting the remaining 
sum over $k_z$ to an integral, we obtain  
\begin{equation}
\label{eq:sigma_zz_intermediate}
    \sigma_{zz}
    =
    \frac{e^2\hbar}{2\pi^2\ell_B^2}
    \sum_{n,s}
    \int dk_z\;
    |v_{z;n,s}(k_z)|^2\,
    \tau_{n,s}^{\mathrm{tr}}(k_z)\,
    \left[
        -\frac{\partial f(E)}{\partial E}
    \right]_{E=E_{n}^s(k_z)} .
\end{equation}
It is important to distinguish between the quantum (single-particle) lifetime
$\tau^{\mathrm{q}}$ and the transport lifetime $\tau^{\mathrm{tr}}$.
The quantum lifetime, defined through the imaginary part of the retarded
self-energy, controls the spectral broadening of Landau levels and determines
the width of single-particle excitations. In contrast, the transport lifetime
accounts for momentum relaxation and incorporates the angular weighting
$(1-\cos\theta)$ that suppresses forward scattering events.
As a result, $\tau^{\mathrm{tr}}$ governs the decay of electrical current and
is the relevant timescale entering the Kubo formula for the longitudinal
conductivity $\sigma_{zz}$. Throughout this work we therefore replace the total
lifetime by $\tau^{\mathrm{tr}}$, which is evaluated microscopically from
screened Coulomb disorder in Appendix~\ref{app:selfenergy}.

\begin{figure*}[t]
\centering
\includegraphics[width=55.5mm,height=45.5mm]{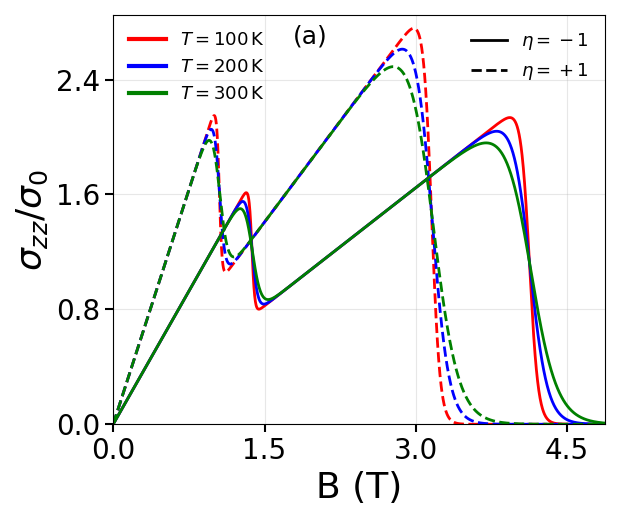}
\includegraphics[width=55.5mm,height=45.5mm]{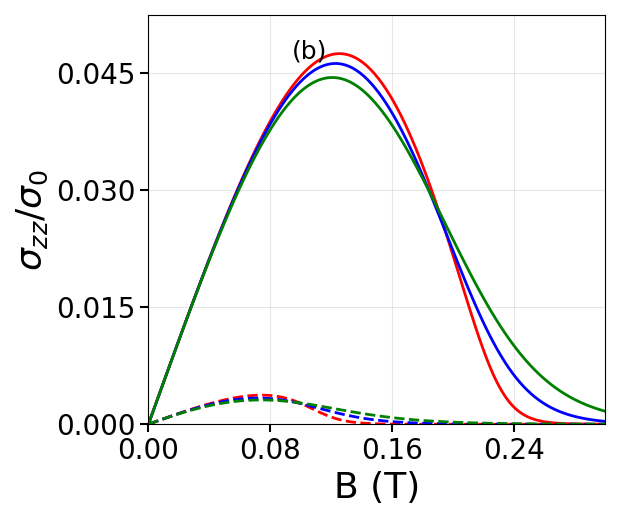}
\includegraphics[width=55.5mm,height=45.5mm]{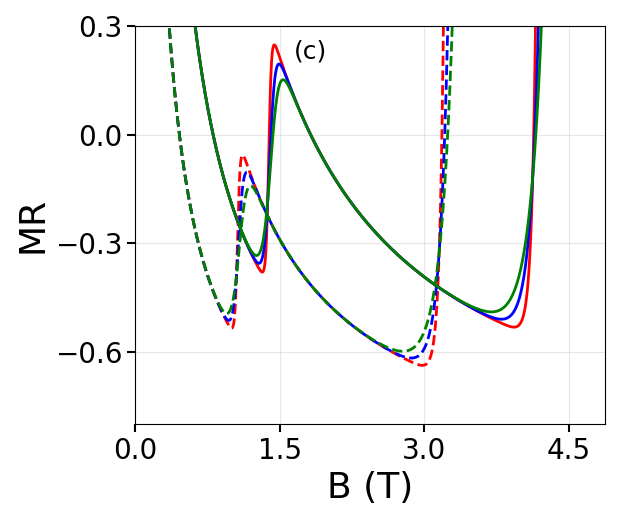}\\
\includegraphics[width=55.5mm,height=45.5mm]{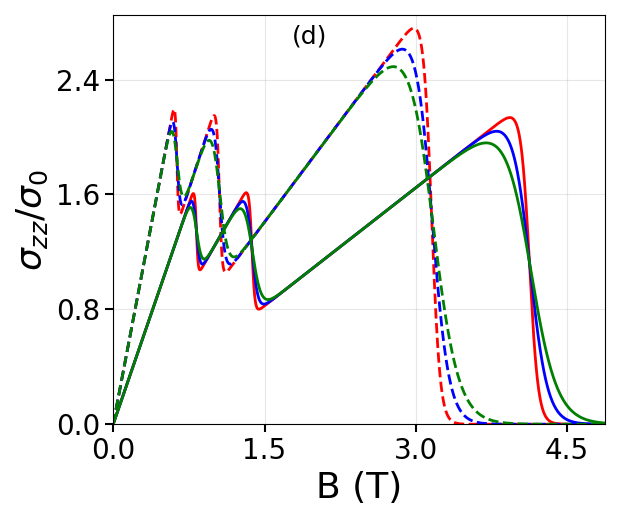}
\includegraphics[width=55.5mm,height=45.5mm]{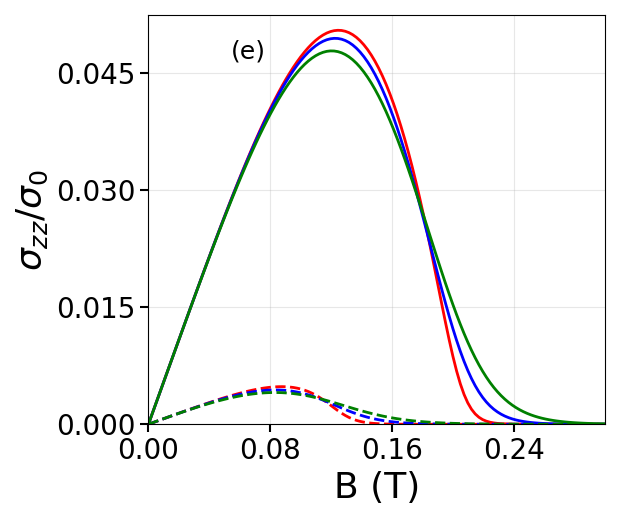}
\includegraphics[width=55.5mm,height=45.5mm]{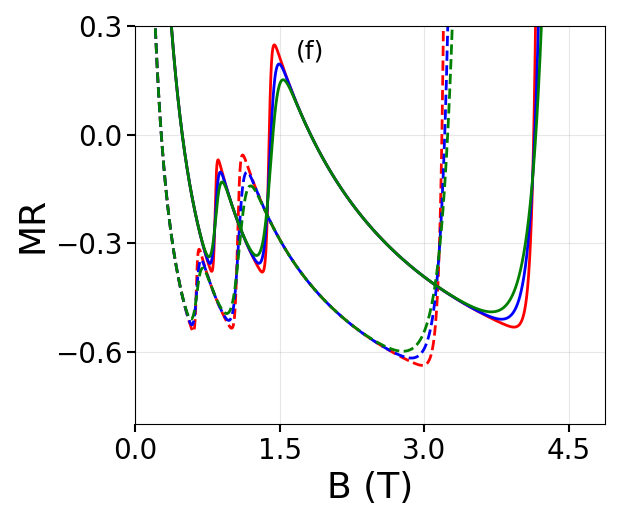}
\caption{(Color online) 
Magnetotransport response of multi-Weyl semimetals for the double-Weyl ($m=2$)
and triple-Weyl ($m=3$) cases.
Red, blue, and green curves correspond to $T=100$, $200$, and $300\,\mathrm{K}$,
respectively.
For $m=2$, two chiral channels give rise to a two-stage linear increase in
$\sigma_{zz}(B)$, while the bulk contribution (b) is confined to low fields,
resulting in negative magnetoresistance (MR).
For $m=3$, three chiral channels produce three linear regimes in
$\sigma_{zz}(B)$, with the bulk contribution (e) again negligible beyond
small fields and a strongly negative MR.
}

\label{fig:m3}
\end{figure*}

\subsection{Contribution from impurity-induced self-scattering of chiral states:}
As stated earlier, the chiral Landau levels, present for $0\le n\le m-1$, exhibit strictly linear dispersion in $k_z$ (Eq. [\ref{eq:tilted_chiral}]) and are topologically protected\cite{nielsen1983adlerbell, BurkovBalents2011}. Using the chiral transport relaxation time derived in Eq. [\ref{chiral_Tr}], the above Eq.~\eqref{eq:sigma_zz_intermediate} becomes
\begin{eqnarray}
\begin{aligned}
&\mathrm{Tr}\!\left[v_z {G_n^\eta}^R(E)\, v_z {G_{n^\prime}^\eta}^A(E)\right]\\
&=2\pi |w_z-\eta v_z|^2 
  \sum_{n=0}^{m-1}
  \mathcal{D}_{\rm ch}^{\eta}(E)\frac{\tau_0}{\mathcal{I}_n^{\eta,\rm ch} }.
\end{aligned}
\end{eqnarray}
where $\mathcal{D}_{\rm ch}^{\eta}(E)$ is the DOS for the chiral states and $\mathcal{I}_n^{\eta,\rm ch}$ is the dimensionless angular integral calculated in the Eq. [\ref{chiral_Tr_1}] of the Appendix with \(\tau_0\) also defined in the same appendix. At $T=0$ K, the derivative of the Fermi distribution becomes a delta function, and Eq. [\ref{generic_s_zz}] thus yields
\begin{equation}\label{sig:chiral}
    \sigma_{zz}^\eta
    =\sigma_0 B
    \sum_{n=0}^{m-1}
    \frac{
    \Theta(E_F-E_n^{\eta,\rm min})\,\Theta(E_n^{\eta,\rm max}-E_F)
    }{\mathcal{I}_n^{\eta,\rm ch}},
\end{equation}
with
\(
    \sigma_0={e^2 |v_{z}| \tau_0}/{l_B^2}
\). 
The steps associated with multiple chiral channels become broadened, but the linear-in-$B$ 
trend persists.


\subsection{Bulk self-scattering and the dominance of the $n=m$ Landau level}

\noindent Using the Eq. \eqref{tau_bulk_app} and the velocity matrix element for the bulk state \(
|v_{z;n,s}^\eta(k_z)|^2
=\left[(w_z+\eta v_z)|\alpha_n^{\eta,s}(k_z)|^2+(w_z-\eta v_z)|\beta_n^{\eta,s}(k_z)|^2\right]^2
\), the longitudinal conductivity can be evaluated from Eq. \eqref{eq:sigma_zz_intermediate}. 
This incorporates the full $k_{z}$ dependence of the 
velocity matrix element, wavefunction amplitudes, and impurity-induced 
scattering rate. 
Although bulk levels with $n>m$ formally appear in the Kubo sum, they may be 
safely omitted when computing $\sigma_{zz}(B)$ for the parameter regime of 
interest. The primary reason is energetic: for the chosen Fermi level 
($E_{F}=200$~meV), all bulk Landau levels with $n>m$ lie far above or below the Fermi level for 
any field in the regime ($B\gtrsim 0.3$-$0.5$~T) and therefore remain 
unoccupied throughout the range where negative magnetoresistance is observed. 
Their exclusion thus does not remove any physically accessible transport 
channel.

The following auxiliary considerations further support this simplification.  
(i) Higher bulk Landau levels possess increasingly oscillatory transverse 
wavefunctions, which strongly enhance their overlap with disorder and produce 
rapidly growing scattering rates and correspondingly short transport lifetimes.  
(ii) Even the lowest bulk level contributes only a small, short-lived 
background at very low fields, while including levels with $n>m$ alters 
$\sigma_{zz}(B)$ by less than $10^{-4}$ yet substantially increases numerical 
cost.  
(iii) All experimentally relevant features, the growth of the $m$ chiral 
channels, the multi-segment structure of $\sigma_{zz}(B)$ for $m>1$, and the 
resulting negative magnetoresistance are governed exclusively by the $m$ 
chiral Landau levels ($n<m$) together with the lowest bulk level ($n=m$).  
Higher bulk levels introduce no new qualitative behavior and make no 
observable quantitative contribution.

For these reasons, retaining only the $n=m$ bulk Landau level captures all 
physically relevant contributions from the bulk states while remaining fully consistent with the 
quantum mechanics of multi-Weyl fermions.

\subsection{Negative Magnetoresistance}

The magnetoresistance (MR) in the $\mathbf{E}\parallel\mathbf{B}$ configuration is defined as
\begin{equation}
\mathrm{MR}(B)
=
\frac{\rho_{zz}(B)-\rho_{zz}(0)}{\rho_{zz}(0)}
=
\frac{\sigma_{zz}(0)}{\sigma_{zz}(B)}-1,
\end{equation}
where $\rho_{zz}=1/\sigma_{zz}$ and the longitudinal conductivity $\sigma_{zz}(B)$ is obtained by 
combining the chiral and bulk contributions. 
A negative value of MR signals an enhancement of $\sigma_{zz}$ with increasing field-an effect that 
arises microscopically from the chiral anomaly and the field-enhanced degeneracy of the chiral 
Landau levels.
Figure~\ref{fig:m3} summarizes the full magnetotransport response of multi-Weyl semimetals.  
For the double-Weyl case ($m=2$), Fig.~\ref{fig:m3}(a) shows that the two chiral Landau levels 
generate two distinct linear regimes in $\sigma_{zz}(B)$, each associated with one chiral branch 
remaining at the Fermi level.  
The bulk background in Fig.~\ref{fig:m3}(b), contributes only 
at small $B$ because higher-order Landau level wavefunctions couple strongly to disorder and thus 
exhibit short transport lifetimes.  
Consequently, the total conductivity is dominated by the chiral states over 
most of the field range, producing a two-stage rise that directly translates into the negative MR 
curve of Fig.~\ref{fig:m3}(d).  This behavior reflects the progressive activation of chiral channels as the magnetic field grows, a 
feature absent in semiclassical descriptions~\cite{son2013chiral,kim2013dirac}.

The triple-Weyl case ($m=3$) further accentuates these trends.  
Each of the three chiral levels contributes a well-defined linear segment in Fig.~\ref{fig:m3}(d), 
leading to a characteristic three-step profile in the total conductivity.  
The bulk background shown in Fig.~\ref{fig:m3}(e) again decays rapidly, reinforcing that bulk modes play a negligible role in the anomaly-driven regime.  
As a result, the MR curve in Fig.~\ref{fig:m3}(f) is strongly negative and displays pronounced 
kinks marking the field values at which individual chiral branches are lifted above the Fermi 
energy. Such multi-step signatures are a direct fingerprint of the node’s monopole charge and cannot be 
captured within semiclassical Boltzmann theories~\cite{son2013chiral,kim2013dirac}.

\begin{figure}[t]
\includegraphics[width=80mm,height=55.5mm]{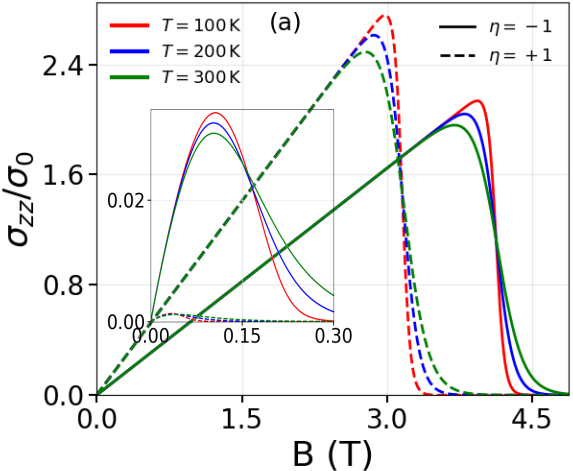}\\
\includegraphics[width=80.5mm,height=55.5mm]{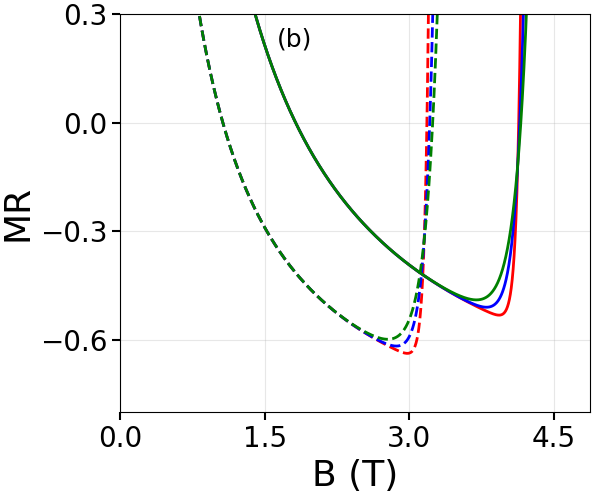}
\caption{
(Color online)
Magnetotransport response of a single-Weyl node ($m=1$).
Panel (a) shows the magnetic-field dependence of the normalized longitudinal
conductivity $\sigma_{zz}/\sigma_0$, dominated by the chiral channel.
The inset highlights the contribution from non-chiral (bulk) (\(n=m\)) Landau level, 
which is confined to the low-field regime and remains negligibly small.
Panel (b) shows the corresponding magnetoresistance (MR), which is negative
over the entire field range shown.
}
\label{fig:m1}
\end{figure}

A convenient way to quantify the multi-step structure of the longitudinal 
conductivity is to identify the magnetic-field values at which each chiral 
Landau level crosses the Fermi energy. For a chiral branch indexed by 
$n$ ($0 \le n \le m-1$), the condition $E_n^{\rm ch}(k_z=0)=E_F$ yields the 
characteristic field scale
\begin{equation}
B_n
=
\frac{E_F + (v_z - \eta w_z)\,\mathcal{Q}/2}{(2n+1)\,w_{\parallel}},
\label{eq:B_n}
\end{equation}
where the factor $(2n+1)$ reflects the equally spaced offsets of the $m$ 
chiral levels. These $B_n$ values correspond directly to the kink locations 
in $\sigma_{zz}(B)$: once $B$ exceeds $B_n$, the $n^{\rm th}$ chiral channel is 
pushed above the Fermi level and no longer contributes to transport. 
In double- and triple-Weyl systems, the resulting sequence 
$\{B_0,B_1,\dots,B_{m-1}\}$ produces the two- and three-stage linear increase of 
$\sigma_{zz}(B)$ observed in Figs.~\ref{fig:m3}(a) \& (d), and the associated 
features in the MR curves in Figs.~\ref{fig:m3}(c) \& (f). 
Thus, the piecewise linear structure of $\sigma_{zz}(B)$ provides a direct 
transport fingerprint of the node’s topological charge through the number 
of anomaly-active chiral modes.

Overall, Fig.~\ref{fig:m3} demonstrates that negative magnetoresistance in multi-Weyl semimetals 
arises from the combined effects of Landau quantization, the fixed number of chiral branches, and 
their field-independent scattering times.  
The linear-in-$B$ growth of $\sigma_{zz}(B)$, the piecewise structure reflecting the monopole 
charge, and the rapid suppression of bulk contributions together provide a clear and robust quantum 
mechanical origin for the anomaly-driven MR response in Weyl and multi-Weyl systems.

As a consistency check, Fig.~\ref{fig:m1} shows the magnetotransport response for
the single-Weyl case ($m=1$). The chiral channel in Fig.~\ref{fig:m1}(a)
produces a single linear rise of $\sigma_{zz}(B)$, while the bulk contribution in the inset of 
Fig.~\ref{fig:m1}(a) appears only at very small fields and rapidly vanishes due
to strong impurity scattering. Consequently, the total conductivity is dominated entirely by the chiral level, yielding the
monotonic negative MR shown in Fig.~\ref{fig:m1}(b). This contrasts with the
multi-step structure observed for $m=2$ and $m=3$, confirming that the richer
NMR profiles originate from the additional chiral channels of higher-charge
nodes.

\section{Conclusion}

We have developed a fully quantum-mechanical theory of negative magnetoresistance in Weyl and
multi-Weyl semimetals based on their Landau-quantized spectrum and microscopic impurity
scattering. Starting from the multi-Weyl Hamiltonian, we showed that the topological charge $m$
determines the number of chiral Landau levels, each providing a unidirectional transport channel
in the $\mathbf{E}\parallel \mathbf{B}$ configuration. Using screened Coulomb disorder within
Fermi’s golden rule and incorporating the resulting relaxation times into the Kubo formalism, we
obtained closed-form expressions for the longitudinal conductivity that capture the distinct roles
of chiral and bulk Landau levels.

Our results demonstrate that the anomaly-driven magnetoconductivity is governed almost entirely
by the chiral modes, whose field-independent scattering time and field-enhanced degeneracy 
produce a robust linear increase of $\sigma_{zz}(B)$. The number of linear segments directly 
reflects the monopole charge: single-, double-, and triple-Weyl nodes exhibit one, two, and three
distinct slopes, respectively. Bulk Landau levels, by contrast, contribute only at very small
fields due to their strong disorder sensitivity and rapidly decaying lifetimes. As a result, the
magnetoresistance is strongly negative and displays characteristic kinks marking the depopulation
of individual chiral branches.

This work provides a unified quantum description of chiral-anomaly-induced transport that goes
beyond semiclassical formulations and makes clear predictions for experimental signatures in
multi-Weyl systems. The multi-step NMR structure, the dominance of the chiral channels, and the
rapid suppression of bulk contributions together offer a direct probe of the topological charge
and Landau-level hierarchy in these materials. Our framework can be readily extended to include
finite-tilt effects, interactions, and more realistic disorder profiles, opening the door to
comprehensive quantum transport modeling of higher-order topological semimetals.

\textit{Acknowledgments}: This work is an outcome of the Research work carried out under the SRG Project, SRG/2023/001516, 
Anusandhan National Research Foundation (ANRF),
Government of India.


\appendix
\section{Impurity-limited self-energy}
\label{app:selfenergy}

\begin{figure*}[http!]
\begin{center}

\begin{tikzpicture}[xscale=0.75, yscale=1.05, thick,
  fermion/.style={thick, draw=blue!70!black, -{Latex[length=4pt,width=4pt]}},
  interaction/.style={decorate, decoration={snake,amplitude=1.6pt,segment length=8pt}, draw=red!80!black, thick}
]

\node at (-10,-0.5) {$\Sigma^{(2)} = $};

\begin{scope}[shift={(-7,0)}, scale=1.1]

\filldraw[fill=gray!45, draw=gray!60, opacity=0.70]
    (1.3,-0.75) ellipse (2.5 and 1.5);

\coordinate (Ltop1) at (-0.3,0.9);
\coordinate (Lbot1) at (-0.3,0.0);
\coordinate (Rtop1) at (2.8,0.9);
\coordinate (Rbot1) at (2.8,0.0);

\foreach \p in {Ltop1,Rtop1}
  \fill (\p) circle (3pt);

\foreach \p in {Lbot1,Rbot1}
  \draw[black, line width=1pt] (\p) circle (3pt);

\draw[fermion] (Lbot1) .. controls (-0.8,0.45) .. (Ltop1);
\draw[fermion] (Ltop1) .. controls (0.2,0.45) .. (Lbot1);

\draw[->,blue!80!black] (0.08,0.45) -- (0.08,0.4);
\draw[->,blue!80!black] (-0.68,0.4) -- (-0.68,0.45);

\draw[fermion] (Rtop1) .. controls (3.3,0.45) .. (Rbot1);
\draw[fermion] (Rbot1) .. controls (2.3,0.45) .. (Rtop1);

\draw[->,blue!80!black] (3.2,0.4) -- (3.2,0.45);
\draw[->,blue!80!black] (2.45,0.45) -- (2.45,0.4);

\draw[fermion, black, ->] (1.2,-0.6) -- (-0.95,-1.35)
      node[midway, below] {$k_F$};


\draw[decorate,decoration={snake,amplitude=1mm,segment length=3mm},
      green!70!black,thick]
  (Rtop1) .. controls (1.4,1.3) and (0.5,1.3) .. (Ltop1);
\draw[->,green!70!black,thick] (1.1,1.42) -- (0.9,1.42);

\draw[decorate,decoration={snake,amplitude=1mm,segment length=3mm},
      green!70!black,thick]
  (Rbot1) .. controls (1.4,-0.4) and (0.5,-0.4) .. (Lbot1);
\draw[->,green!70!black,thick] (0.9,-0.52) -- (1.1,-0.52);

\node at (3.5,-0.2) {\small \(|n,k_z\rangle\)};
\node at (4.2,0.9) {\small \(|n' ,k_z-q_z\rangle\)};
\node at (-1.8,0.9) {\small \(|n ,k_z'+q_z\rangle\)};
\node at (-0.9,-0.2) {\small \(|n',k_z'\rangle\)};

\node at (1.2,-1.3) {\small (a)};

\end{scope}

\node at (-1.8,-0.5) {\Large $+$};

\begin{scope}[shift={(1.5,0)}, scale=1.1]

\filldraw[fill=gray!45, draw=gray!60, opacity=0.70]
    (1.3,-0.75) ellipse (2.5 and 1.5);

\coordinate (Ltop2) at (-0.3,0.9);
\coordinate (Lbot2) at (-0.3,0.0);
\coordinate (Rtop2) at (2.8,0.9);
\coordinate (Rbot2) at (2.8,0.0);

\foreach \p in {Ltop2,Rtop2}
  \fill (\p) circle (3pt);

\foreach \p in {Lbot2,Rbot2}
  \draw[black, line width=1pt] (\p) circle (3pt);

\draw[fermion] (Lbot2) .. controls (-0.8,0.45) .. (Ltop2);
\draw[fermion] (Rtop2) .. controls (3.3,0.45) .. (Rbot2);

\draw[fermion] (Ltop2) -- (Rbot2);
\draw[fermion] (Lbot2) -- (Rtop2);

\draw[->,blue!80!black] (-0.68,0.4) -- (-0.68,0.45);
\draw[->,blue!80!black] (3.2,0.4) -- (3.2,0.45);
\draw[->,blue!80!black] (2.2,0.17) -- (2.3,0.12);
\draw[->,blue!80!black] (0.3,0.17) -- (0.15,0.12);

\draw[decorate,decoration={snake,amplitude=1mm,segment length=3mm},
      green!70!black,thick]
  (Rtop2) .. controls (1.4,1.3) and (0.5,1.3) .. (Ltop2);
\draw[->,green!70!black,thick] (1.1,1.42) -- (0.9,1.42);

\draw[decorate,decoration={snake,amplitude=1mm,segment length=3mm},
      green!70!black,thick]
  (Rbot2) .. controls (1.4,-0.4) and (0.5,-0.4) .. (Lbot2);
\draw[->,green!70!black,thick] (0.9,-0.52) -- (1.1,-0.52);

\node at (1.2,-1.3) {\small (b)};
\draw[fermion, black, ->] (1.2,-0.6) -- (-0.95,-1.35)
      node[midway, below] {$k_F$};
\node at (3.5,-0.2) {\small \(|n,k_z\rangle\)};
\node at (4.1,0.9) {\small \(|n' ,k_z-q_z\rangle\)};
\node at (-1.4,0.9) {\small \(|n ,k_z'+q_z\rangle\)};
\node at (-0.9,-0.2) {\small \(|n',k_z'\rangle\)};

\end{scope}

\end{tikzpicture}

\caption{
Second-order impurity self-energy diagrams:  
(a) direct (Hartree) contribution, giving only a real energy shift;  
(b) exchange (crossed) contribution, producing the imaginary part of
$\Sigma^{R}$ and thus the impurity scattering rate.
}

\label{fig:diagrams}
\end{center}
\end{figure*}
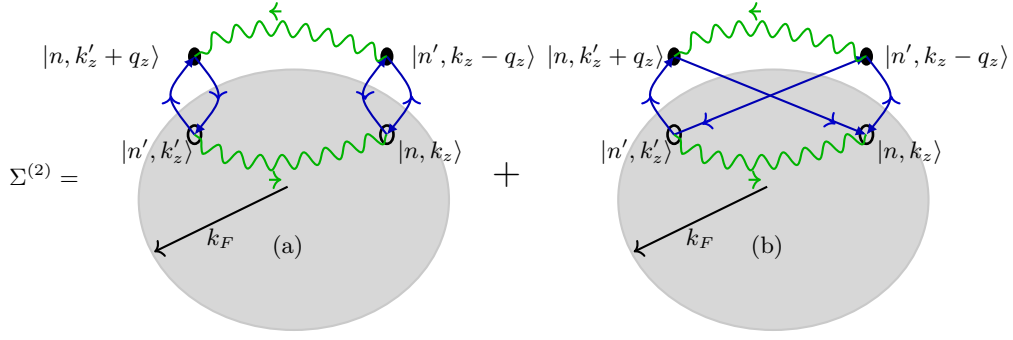

In this Appendix we derive the impurity-induced self-energy in the Landau-level
basis $\ket{n,k_z}$ used throughout the main text.
In the presence of a magnetic field, the impurity vertices and momentum routing
are constrained by Landau quantization, which is encoded diagrammatically in
Fig.~\ref{fig:diagrams}.
These diagrams determine the structure of the Dyson equation and directly fix
the form of the quantum and transport scattering rates derived below.

We consider static screened Coulomb impurities located at random positions
$\{\mathbf R_j\}$.
The total impurity potential is
$U(\mathbf r)=\sum_j \mathcal{V}(\mathbf r-\mathbf R_j)$, with Fourier transform
$U(\mathbf q)=\mathcal{V}(\mathbf q)\sum_j e^{-i\mathbf q\cdot\mathbf R_j}$.
For uncorrelated impurities of density $n_i$, disorder averaging yields
\begin{equation}
\langle U(\mathbf q)U(\mathbf q')\rangle_{\rm dis}
=
n_i |\mathcal{V}(\mathbf q)|^2 (2\pi)^3\delta(\mathbf q+\mathbf q'),
\label{eq:imp-cor-appendix}
\end{equation}
which enforces the momentum pairing $(\mathbf q,-\mathbf q)$ along each impurity
line in Fig.~\ref{fig:diagrams}.

\paragraph*{Screened Coulomb disorder:}
We model the impurity potential by a statically screened Coulomb interaction,
\begin{equation}
\mathcal{V}(\mathbf q)
=
\frac{4\pi e^2}{\varepsilon(q^2+\kappa^2)},
\label{eq:Vq}
\end{equation}
where $\varepsilon$ is the background dielectric constant and $\kappa$ is the
Thomas-Fermi screening wave vector.
For a finite carrier density, $\kappa$ follows from the static Lindhard
expression,
\begin{equation}
\kappa^2
=
\frac{4\pi e^2}{\varepsilon}
\left.\frac{\partial n_e}{\partial\mu}\right|_{\mu=E_F}
=
\frac{4\pi e^2}{\varepsilon}\mathcal D(E_F).
\label{eq:kappa}
\end{equation}

For a representative Fermi energy $E_F=200$~meV and magnetic fields relevant to
this work, the Fermi level lies within the energy windows of the anomaly-active
chiral Landau levels up to their depopulation fields.
In this regime all $m$ chiral branches contribute, reducing Eq. \eqref{Dos_Chi} to
\begin{equation}
\mathcal D_{\rm ch}(E_F)
\simeq
\frac{m}{4\pi^2 l_B^2 |w_z-\eta v_z|}.
\label{eq:DOS_chiral_EF}
\end{equation}
Substitution into Eq.~\eqref{eq:kappa} yields
\begin{equation}
\kappa^2
=
\frac{e^3 m B}{\pi\varepsilon\hbar |w_z-\eta v_z|},
\label{eq:kappa_TF_final}
\end{equation}
so that the Thomas-Fermi screening length scales as
\begin{equation}
\lambda_{\rm TF}=\kappa^{-1}\propto(mB)^{-1/2}.
\end{equation}
Using representative material parameters, we estimate
$\lambda_{\rm TF}\sim1$-$3$~nm for $B\sim1$-$4$~T, confirming that Coulomb
disorder is efficiently screened in the anomaly-dominated regime.

\paragraph*{Impurity vertex and Dyson equation:}
In a magnetic field the eigenstates factorize as
$\ket{n,k_z}=\ket{n}_\perp\otimes\ket{k_z}$.
The impurity vertex therefore takes the form
\begin{equation}
\langle n,k_z|e^{i\mathbf q\cdot\mathbf r}|n',k_z'\rangle
=
\delta_{k_z',\,k_z-q_z}\,\mathcal F_{nn'}(\mathbf q_\perp),
\label{eq:vertex-appendix}
\end{equation}
which enforces longitudinal momentum transfer $q_z$ and Landau-level mixing
through the form factor $\mathcal F_{nn'}$.
These constraints are represented graphically in Fig.~\ref{fig:diagrams}.

The disorder-averaged Green’s function satisfies the Dyson equation
\begin{equation}
G=G_0+G_0\Sigma G,
\end{equation}
which generates the two second-order diagrams shown in Fig.~\ref{fig:diagrams}.
The direct (Hartree) diagram produces only a real, momentum-independent energy
shift and does not contribute to level broadening or transport.
It will therefore not be considered further.

The exchange diagram, Fig.~\ref{fig:diagrams}(b), contains a genuine momentum
exchange between the two fermionic trajectories.
Evaluating the vertices using Eq.~\eqref{eq:vertex-appendix}, one obtains the
retarded self-energy
\begin{equation}
\Sigma_n^R(k_z,\omega)=
\frac{N_i}{V}
\sum_{n'}\sum_{\mathbf q}
|\mathcal V(\mathbf q)|^2
|\mathcal F_{nn'}(\mathbf q_\perp)|^2
G_{n'}^R(k_z-q_z,\omega),
\label{eq:SE-final-appendix}
\end{equation}
which directly corresponds to diagram~(b).

\paragraph*{Quantum and transport lifetimes.}
The imaginary part of Eq.~\eqref{eq:SE-final-appendix}, evaluated on shell,
determines the quantum lifetime,
\begin{eqnarray}
\frac{1}{\tau_n^{\rm q}(k_z)}
&=
2\pi\frac{N_i}{V}
\sum_{n'}\sum_{\mathbf q}
|\mathcal V(\mathbf q)|^2
|\mathcal F_{nn'}(\mathbf q_\perp)|^2
\nonumber\\&\times \delta\!\left[
\varepsilon_n(k_z)-\varepsilon_{n'}(k_z-q_z)
\right].
\label{eq:tau-q-appendix}
\end{eqnarray}
The transport lifetime is obtained by including the usual factor
$(1-\cos\theta)$, which suppresses forward scattering,
\begin{eqnarray}\label{inv_transport_time}
\frac{1}{\tau_n^{\rm tr}(k_z)}
&=
2\pi\frac{N_i}{V}
\sum_{n'}\sum_{\mathbf q}
|\mathcal V(\mathbf q)|^2
|\mathcal F_{nn'}(\mathbf q_\perp)|^2
(1-\cos\theta)
\nonumber\\&\times \delta\!\left[
\varepsilon_n(k_z)-\varepsilon_{n'}(k_z-q_z)
\right].
\label{eq:tau-tr-appendix}
\end{eqnarray}
Only $\tau_n^{\rm tr}$ enters the Kubo conductivity.

\paragraph*{Validity of the Born approximation:}
The second-order Born expansion is controlled provided the disorder-induced
broadening $\Gamma_n=-\Im\Sigma_n^R$ remains small compared to the Landau-level
spacing.
For chiral levels this spacing is
$\Delta_n=E_{n+1}^{\rm ch}-E_n^{\rm ch}=2Bw_\parallel$,
and we require $\Gamma_n\ll\Delta_n$.
Using the screened potential~\eqref{eq:Vq}, we find that $\Gamma_n$ decreases with
field for chiral modes and grows only weakly for bulk levels due to the rapid
decay of the form factors.
For impurity densities $n_{\rm imp}=10^{16}$-$10^{18}\,\mathrm{cm}^{-3}$, this
condition is satisfied for all fields $B\gtrsim0.5$~T.


\subsection{\textbf{\textit{Derivation of the form factor }} \(\mathcal{F}_{nn^\prime}({\bf q_\perp})\):}
\noindent\textbf{\textit{For Chiral states }}: The Landau-level eigenstate for the chiral states 
takes the spinor of the form
\[
\ket{\psi_{n}^{\rm ch}}
=
\begin{pmatrix}
0\\[4pt]
\ket{n}
\end{pmatrix}
\otimes \ket{k_{x},k_{z}}.
\]

\noindent To obtain the position-space representation, we use the standard relations
\(
\langle x|k_{x}\rangle={e^{ik_{x}x}}/{\sqrt{L_{x}}}, \qquad
\langle z|k_{z}\rangle={e^{ik_{z}z}}/{\sqrt{L_{z}}},
\)
and
\(
\langle y|n\rangle = {c_{n}}/{\sqrt{l_{B}}}
\, e^{-\zeta^{2}/2} \mathcal{H}_{n}(\zeta), \qquad \text{with }
c_{n}={1}/{\sqrt{2^{\,n} n!\sqrt{\pi}}}
\) and \(\mathcal{H}_n(\zeta)\) being the Hermite polynomial of degree \(n\).

Putting these results together, the chiral state's real-space wavefunction, written in Schroedinger notation, takes the form
\begin{eqnarray}\label{psi_space}
\psi_{n}^{\rm ch}(\mathbf{r})=
\frac{c_{n}}{\sqrt{L_{x}L_{z}l_{B}}}\,
e^{ik_{x}x}\, e^{ik_{z}z}\,
e^{-\frac{(y-y_{0})^{2}}{2l_{B}^{2}}}\,
\mathcal{H}_{n}\!\left(\frac{y-y_{0}}{l_{B}}\right),
\end{eqnarray}
where we have defined \(y_0=k_x l_B^2\).

Using the explicit form of the wavefunctions, the matrix element becomes
\begin{widetext}
    \begin{eqnarray}
\langle \psi_{n}^{\rm ch}|U|\psi_{n^\prime}^{\rm ch}\rangle
&=& \sum_{i}\int \frac{d^{3}r}{L_xL_z}\,
     \psi_{n'}^{*}(y) \mathcal{V}(\mathbf{r}-\mathbf{R}_i)
     \psi_{n}(y)
     e^{i(k_x-k_x')x}
     e^{i(k_z-k_z')z}
\nonumber\\
 &=&
\sum_{i}\,
(2\pi\delta_{q_x,k_x-k_x^\prime})
(2\pi\delta_{q_z,k_z-k_z^\prime})
\int \frac{d^{3}q}{(2\pi)^{3}}
    \mathcal{V}(\mathbf{q}) e^{-i\mathbf{q}\cdot\mathbf{R}_i}
\int dy\, \psi_{n^\prime}^{*}(y)\psi_{n}(y)e^{iq_y y}.
\label{eq:matrix_element_chiral}
\end{eqnarray}
We separate out the non-trivial \(y\)-integral,
\begin{eqnarray}
I_y^c=\frac{c_nc_{n^\prime}}{l_B}
\int_{-\infty}^{\infty}
e^{-(y-y_0)^2/2l_B^2}
e^{-(y-y_0^\prime)^2/2l_B^2}
\mathcal{H}_n\!\left(\frac{y-y_0}{l_B}\right)
\mathcal{H}_{n^\prime}\!\left(\frac{y-y_0^\prime}{l_B}\right)
e^{iq_yy}\,dy.
\end{eqnarray}

Using the exact Hermite-polynomial identity
\begin{equation}
\int_{-\infty}^{\infty} e^{-x^2}
\mathcal{H}_n(x+a)\mathcal{H}_m(x+b)\,dx 
=2^n\sqrt{\pi}\,n!(b-a)^{|m-n|}
\mathcal{L}_n^{|m-n|}(-2ab)
\end{equation}
we find an analytical expression involving associated Laguerre polynomials. A more compact form uses
\(
q_{\perp}=\frac{l_B^{2}}{2}(q_x^{2}+q_y^{2}),
\)
leading to
\begin{equation}
I_y^c =
\sqrt{\frac{n_<!}{n_>!}}\,
e^{iq_y(y_0+y_0^\prime)/2}
e^{-q_{\perp}/2}
\left[\frac{l_B}{2}(q_x+iq_y)\right]^{|n-n^\prime|}
\mathcal{L}_{n_<}^{|n-n^\prime|}(q_{\perp}).
\end{equation}

For chiral self-scattering (i.e., \(n=n^\prime\)), this simplifies to
\(
I_y=e^{iq_y(y_0+y_0^\prime)/2}e^{-q_{\perp}/2}\mathcal{L}_n(q_{\perp}),
\)
thus giving us 

\begin{eqnarray*}
\langle \psi_{n}^{\rm ch}|U|\psi_{n^\prime}^{\rm ch}\rangle
=
\sum_{i}\,
(2\pi)^2\delta_{q_x,k_x-k_x^\prime}
\delta_{q_z,k_z-k_z^\prime} \sqrt{\frac{n_<!}{n_>!}}\,
e^{iq_y(y_0+y_0^\prime)/2}
e^{-q_{\perp}/2} 
\left[\frac{l_B}{2}(q_x+iq_y)\right]^{|n-n^\prime|}
\mathcal{L}_{n_<}^{|n-n^\prime|}(q_{\perp})\int \frac{d^{3}q}{(2\pi)^{3}}
    \mathcal{V}(\mathbf{q}) e^{-i\mathbf{q}\cdot\mathbf{R}_i}
\label{eq:matrix_element_chiral}
\end{eqnarray*}
We can write  \\
\begin{eqnarray}
    \mathcal{F}_{nn^\prime}({\bf q_\perp})=\sqrt{\frac{n_<!}{n_>!}}\,
e^{iq_y(y_0+y_0^\prime)/2}
e^{-q_{\perp}/2} \left[\frac{l_B}{2}(q_x+iq_y)\right]^{|n-n^\prime|}
\mathcal{L}_{n_<}^{|n-n^\prime|}(q_{\perp})
\end{eqnarray}

\noindent Using the above results in Eq. ~\ref{inv_transport_time}, we obtained the inverse relaxation time states \({1}/{\tau^{\mathrm{tr}}}\) as 
\begin{equation}
\frac{1}{\tau^{\mathrm{tr}}}
=
\frac{n_{\mathrm{imp}}}{2k_F^{2}}
\frac{1}{|w_z-\eta v_z|}
\int_{0}^{\infty}dq_{\perp}\,
\frac{q_{\perp}\,e^{-q_{\perp}}\,[\mathcal{L}_n(q_{\perp})]^{2}}
{\left(q_{\perp}+\frac{1}{4\pi^{2}|w_z-\eta v_z|}\right)^{2}},
\end{equation}
which can be expressed in terms of \(\mathcal{I}_n^{\eta,\rm ch}\) as
\begin{equation}\label{chiral_Tr}
\tau^{\mathrm{tr}} = \frac{\tau_{0}}{\mathcal{I}_n^{\eta,\rm ch}},
\end{equation}
with \(\tau_0=\frac{2k_F^{2} V\epsilon |w_z-\eta v_z|}{n_{\mathrm{imp}}}\) and we have defined
\begin{equation}\label{chiral_Tr_1}
  \mathcal{I}_n^{\eta,\rm ch}=  \int_{0}^{\infty}dq_{\perp}\,
\frac{q_{\perp}\,e^{-q_{\perp}}\,[\mathcal{L}_n(q_{\perp})]^{2}}
{\left(q_{\perp}+\frac{1}{4\pi^{2}|w_z-\eta v_z|}\right)^{2}}.
\end{equation}

\noindent\textbf{\textit{For Bulk states: }}
We can similarly write the real-space wavefunction (in Schrodinger notation) for a bulk state \(n\ge m\) in conduction band as,
\begin{eqnarray}
    \psi_{n}^b(\mathbf{r})=
\frac{c_{n}}{\sqrt{L_{x}L_{z}l_{B}}}\,
e^{ik_{x}x}\, e^{ik_{z}z}\,
e^{-\frac{(y-y_{0})^{2}}{2l_{B}^{2}}}
[\alpha_n+\beta_n \mathcal{H}_{n}\!\left(\frac{y-y_{0}}{l_{B}}\right)].
\end{eqnarray}
We drop the \(\eta,s\) indices from the wavefunction, noting that the above equation pertains to a bulk state in the conduction band of a particular chirality.
Everything in $ \langle \psi_{n}^b|U|\psi_{n^\prime}^b\rangle$ will remain unchanged except the y-integral, thus yielding the form factor,
\begin{eqnarray}\label{Iy_bulk}
  \mathcal{F}_{nn^\prime}({\bf q_\perp})=\frac{c_nc_{n^\prime}}{l_B}
\int_{-\infty}^{\infty}
e^{-(y-y_0)^2/2l_B^2}
e^{-(y-y_0^\prime)^2/2l_B^2}
[\alpha_n+\beta_n \mathcal{H}_{n}\!\left(\frac{y-y_0}{l_B}\right)]
[\alpha_{n^\prime}+\beta_{n^\prime}\mathcal{H}_{n^\prime}\!\left(\frac{y-y_0^\prime}{l_B}\right)]
e^{iq_yy}\,dy
\end{eqnarray}

\noindent After solving Eq.~\ref{Iy_bulk} by only taking the intra-level scattering $n=n^\prime$, we have,
\begin{equation}
\mathcal{F}_{nn}(\mathbf q_\perp)
=
e^{-q_\perp/2 + i q_y (y_0+y_0^\prime)/2}
\left[
\mathcal{L}_n(q_\perp) + f_n(q_\perp,\phi)
\right].
\end{equation}
where for compactness we define,
\(
f_n(q_\perp,\phi)=
\begin{cases}
\dfrac{2^{-n}\alpha_n\!\left(\alpha_n+2l_B^{n}\beta_n q^{\,n}\cos\phi\right)}{n!},
& n \ \text{even}, \\[10pt]
\dfrac{2^{-n}\alpha_n\!\left(\alpha_n+2l_B^{n}\beta_n q^{\,n}\sin\phi\right)}{n!},
& n \ \text{odd}.
\end{cases}
\)

\noindent Using the above results in Eq. ~\ref{inv_transport_time}, and using the replacement
\(
\sum_{\mathbf q}
\;\rightarrow\;
\frac{V}{(2\pi)^3}
\int dq_z \int_0^\infty \frac{2}{l_B^2} dq_\perp \int_0^{2\pi} d\phi,
\)
we obtain the inverse relaxation time for the bulk state, \(|m,k_z\rangle\), as,
\begin{align}
\frac{1}{\tau_n^{\mathrm{tr}}(k_z)}
=
\frac{2\pi n_i}{(2\pi)^3}
\int dq_z \int_0^\infty \frac{2}{l_B^2} dq_\perp
\int_0^{2\pi} d\phi \,
|\mathcal{V}({\bf q})|^2\,
\frac{\frac{2}{l_B^2}q_\perp+q_z^2}{k_f^2}
e^{-q_\perp}
\left[
\mathcal{L}_n^{2}
+2\mathcal{L}_n f_n
+f_n^{2}
\right]
\delta\!\bigl(
\varepsilon_n(k_z)-\varepsilon_n(k_z-q_z)
\bigr).
\end{align}

\noindent Using the linear expansion
\(
\varepsilon_n(k_z)-\varepsilon_n(k_z-q_z)
\simeq 
\left.\frac{\partial \varepsilon_n}{\partial k_z}\right|_{k_z} q_z
\simeq (v_z+\eta w_z)\, q_z,
\)
we obtain
\(
\delta\!\left((v_z+\eta w_z) q_z\right)
=
\frac{1}{|(v_z+\eta w_z)|}\, \delta(q_z).
\)
Thus 
yielding the final expression
\begin{equation}\label{tau_bulk_app}
\boxed{
\frac{1}{\tau_n^{\mathrm{tr}}(k_z)}
=
\frac{2n_i}{\pi |(v_z+\eta w_z)| k_f^{2}l_B^4}
\int_0^\infty q_\perp e^{-q_\perp} dq_\perp
\int_0^{2\pi} d\phi \;
|\mathcal{V}(q_\perp)|^2
\left[
\mathcal{L}_n^{2}
+ 2\mathcal{L}_n f_n
+ f_n^{2}
\right].
}
\end{equation}

\end{widetext}


\begin{thebibliography}{55}
\bibitem{son2013chiral}
D.~T.~Son and B.~Z.~Spivak,
\textit{Chiral anomaly and classical negative magnetoresistance of Weyl metals},
\href{https://doi.org/10.1103/PhysRevB.88.104412}{Phys.\ Rev.\ B {\bf 88}, 104412 (2013)}.

\bibitem{burkov2015qm}
A.~A.~Burkov,
\textit{Negative longitudinal magnetoresistance in Dirac and Weyl metals},
\href{https://doi.org/10.1103/PhysRevB.91.235157}{Phys.\ Rev.\ B {\bf 91}, 245157 (2015)}.

\bibitem{zrj2015nmr}
C.~Zhang \textit{et al.},
\textit{Observation of the Adler-Bell-Jackiw anomaly in a Weyl semimetal},
\href{https://doi.org/10.1038/ncomms10735}{Nat.\ Commun.\ {\bf 7}, 10735 (2016)}.

\bibitem{kim2013dirac}
H.-J.~Kim \textit{et al.},
\textit{Dirac versus Weyl Fermions in Topological Insulators: Adler-Bell-Jackiw Anomaly
in Transport Phenomena},
\href{https://doi.org/10.1103/PhysRevLett.111.236603}{Phys.\ Rev.\ Lett.\ {\bf 111}, 246603 (2013)}.

\bibitem{huang2015nmr}
X.~Huang \textit{et al.},
\textit{Observation of the chiral anomaly induced negative magnetoresistance in 3D Weyl semimetal TaAs},
\href{https://doi.org/10.1103/PhysRevX.5.031023}{Phys.\ Rev.\ X {\bf 5}, 031023 (2015)}.


\bibitem{spivak2016hydro}
B.~Z.~Spivak and A.~V.~Andreev,
\textit{Magnetotransport phenomena related to the chiral anomaly in Weyl semimetals},
\href{https://doi.org/10.1103/PhysRevB.93.085107}{Phys.\ Rev.\ B {\bf 93}, 085107 (2016)}.

\bibitem{wan2011wsmdirac}
X.~Wan, A.~M.~Turner, A.~Vishwanath, and S.~Y.~Savrasov,
\textit{Topological semimetal and Fermi-arc surface states in the electronic structure of pyrochlore iridates},
\href{https://doi.org/10.1103/PhysRevB.83.205101}{Phys.\ Rev.\ B {\bf 83}, 205101 (2011)}.

\bibitem{burkov2011weylmetal}
A.~A.~Burkov and L.~Balents,
\textit{Weyl semimetal in a topological insulator multilayer},
\href{https://doi.org/10.1103/PhysRevLett.107.127205}{Phys.\ Rev.\ Lett.\ {\bf 107}, 127205 (2011)}.

\bibitem{armitage2018review}
N.~P.~Armitage, E.~J.~Mele, and A.~Vishwanath,
\textit{Weyl and Dirac semimetals in three-dimensional solids},
\href{https://doi.org/10.1103/RevModPhys.90.015001}{Rev.\ Mod.\ Phys.\ {\bf 90}, 015001 (2018)}.

\bibitem{yan2017topological}
B.~Yan and C.~Felser,
\textit{Topological materials: Weyl semimetals},
\href{https://doi.org/10.1146/annurev-conmatphys-031016-025458}{Annu.\ Rev.\ Condens.\ Matter Phys.\ {\bf 8}, 337 (2017)}.

\bibitem{nielsen1983adlerbell}
H.~B.~Nielsen and M.~Ninomiya,
\textit{The Adler-Bell-Jackiw anomaly and Weyl fermions in a crystal},
\href{https://doi.org/10.1016/0370-2693(83)91529-0}{Phys.\ Lett.\ B {\bf 130}, 389 (1983)}.

\bibitem{burkov2014ahe}
A.~A.~Burkov,
\textit{Anomalous Hall Effect in Weyl metals},
\href{https://doi.org/10.1103/PhysRevLett.113.187202}{Phys.\ Rev.\ Lett.\ {\bf 113}, 187202 (2014)}.




\bibitem{fang2012multiweyl}
C. Fang, M. J. Gilbert, X. Dai, and B. A. Bernevig,
"Multi-Weyl Topological Semimetals Stabilized by Point Group Symmetry,"
\href{https://journals.aps.org/prl/abstract/10.1103/PhysRevLett.108.266802}{Phys. Rev. Lett. \textbf{108}, 266802 (2012).}

\bibitem{fang2012multiweyl_1}
C.~Fang, M.~J.~Gilbert, and B.~A.~Bernevig,
\textit{Bulk topological invariants in noninteracting point group symmetric insulators},
\href{https://doi.org/10.1103/PhysRevB.86.115112}{Phys.\ Rev.\ B {\bf 86}, 115112 (2012)}.



\bibitem{Y_Sun}
Y. Sun and A. M. Wang,
Y.~Lai, H.~H.~Lai, and W.~F.~Tsai,
\textit{Magneto-optical conductivity of double Weyl semimetals},
\href{https://journals.aps.org/prb/abstract/10.1103/PhysRevB.96.085147}{Phys.\ Rev.\ B {\bf 96}, 085147 (2017)}.


\bibitem{H_Weng_PRX}
H. Weng, C. Fang, Z. Fang, B. A. Bernevig, and X. Dai,
\textit{Weyl Semimetal Phase in Noncentrosymmetric Transition-Metal Monophosphides},
\href{https://doi.org/10.1103/PhysRevX.5.011029}{Phys.\ Rev.\ X {\bf 5}, 011029 (2015)}.

\bibitem{Y_Xu_prl}
Y. Xu, Z. Song, Z. Wang, H. Weng, and X. Dai,
\textit{Higher-Order Topology of the Axion Insulator EuIn$_2$As$_2$},
\href{https://journals.aps.org/prl/abstract/10.1103/PhysRevLett.122.256402}{Phys.\ Rev.\ Lett {\bf 122}, 255402 (2019)}.

\bibitem{S_Kimura_prb}
S. Kimura, Y. Nakajima, Z. Mita, R. Jha, R. Higashinaka, T. D. Matsuda, Y. Aoki,
\textit{Optical evidence of type-II Weyl semimetals MoTe$_2$ and WTe$_2$},
\href{https://journals.aps.org/prb/abstract/10.1103/PhysRevB.99.195203}{Phys.\ Rev.\ B {\bf 99}, 195203 (2019)}.

\bibitem{xu2017laalge}
Su-Yang Xu \textit{et al.},
\textit{Discovery of Lorentz-violating type II Weyl fermions in LaAlGe},
\href{https://doi.org/10.1126/sciadv.1603266}{Science Advances {\bf 3}, e1603266 (2017)}.

\bibitem{chang2016laalge}
G.~Chang \textit{et al.},
\textit{Unconventional Chiral Fermions and Large Topological Fermi Arcs in RhSi},
\href{https://doi.org/10.1103/PhysRevLett.119.206401}{Phys.\ Rev.\ Lett.\ {\bf 119}, 206401 (2017)}.

\bibitem{liu2018co3sn2s2}
E.~Liu \textit{et al.},
\textit{Giant anomalous Hall effect in a ferromagnetic kagome-lattice semimetal},
\href{https://doi.org/10.1038/s41567-018-0234-5}{Nature Physics {\bf 14}, 1125-1131 (2018)}.

\bibitem{wang2018co3sn2s2}
J. Y. Liu \textit{et al.},
\textit{A magnetic topological semimetal Sr$_{1-y}$Mn$_{1-z}$Sb$_2$ (y,z<0.1)},
\href{https://www.nature.com/articles/nmat4953}{Nature Materials, \textbf{16}, 905-910 (2017)}.

\bibitem{chang2018alpt}
G.~Chang \textit{et al.},
\textit{Topological quantum properties of chiral crystals},
\href{https://doi.org/10.1038/s41563-018-0162-3}{Nature Materials {\bf 17}, 978-
985 (2018)}.

\bibitem{schroter2019topological}
N.~B.~M.~Schroter \textit{et al.},
\textit{Chiral topological semimetal with multifold band crossings and long Fermi arcs},
\href{https://doi.org/10.1038/s41567-019-0511-y}{Nature Physics {\bf 15}, 759-765 (2019)}.

        \bibitem{Lv2015TaAs}
B.~Q.~Lv \textit{et al.},
``Observation of Weyl nodes in TaAs,''
\href{https://www.nature.com/articles/nphys3426}{Nat.\ Phys.\ \textbf{11}, 724 (2015).}

\bibitem{NXu2015TaAs}
S. Y.~Xu \textit{et al.},
``Discovery of a Weyl fermion state with Fermi arcs in niobium arsenide,''
\href{https://www.nature.com/articles/nphys3437}{Nat.\ Phys.\ \textbf{11}, 748-754 (2015).}

\bibitem{Yang2015NbP}
L.~X.~Yang \textit{et al.},
``Weyl semimetal phase in the non-centrosymmetric compound TaAs,''
\href{https://www.nature.com/articles/nphys3425}{Nat.\ Phys.\ \textbf{11}, 728 (2015).}



\bibitem{Huang2016SrSi2}
S.-M.~Huang \textit{et al.},
``New type of Weyl semimetal with quadratic doubleWeyl fermions,''
\href{https://www.pnas.org/doi/epdf/10.1073/pnas.1514581113}{Proc.\ Natl.\ Acad.\ Sci.\ USA \textbf{113}, 1180 (2015).}



\bibitem{Tang2017CoSi}
P.~Tang, Q.~Zhou, and S.-C.~Zhang,
``Multiple types of topological fermions in transition metal silicides,''
\href{https://journals.aps.org/prl/abstract/10.1103/PhysRevLett.119.206402}{Phys.\ Rev.\ Lett.\ \textbf{119}, 206402 (2017).}

\bibitem{Schroter2019RhSi}
N. B. M. Schröter \textit{et al.},
``Topological chiral crystals with multifold band crossings and long Fermi arcs,''
\href{https://www.nature.com/articles/s41567-019-0511-y}{Nat. Phys. \textbf{15}, 759-765 (2019).}


\bibitem{tchoumakov2016tilt}
S.~Tchoumakov, M.~Civelli, and M.~O.~Goerbig,
``Magnetic-field-driven relativistic properties in type-I and type-II Weyl semimetals,''
\href{https://journals.aps.org/prl/abstract/10.1103/PhysRevLett.117.086402}{{Phys. Rev. Lett.} \textbf{117}, 086402 (2016).}

\bibitem{soluyanov2015typeII}
A.~A.~Soluyanov, D.~Gresch, Z.~Wang, Q.~Wu, M.~Troyer, X.~Dai, and B.~A.~Bernevig,
``Type-II Weyl semimetals,'' 
\href{https://www.nature.com/articles/nature15768}{{Nature} \textbf{527}, 495-498 (2015).}




\bibitem{s_saha} 
Sushmita Saha, Deepannita Das, Alestin Mawrie, 
\textit{Unveiling the Chiral States in Multi-Weyl Semimetals through Magneto-Optical Spectroscopy}, 
\href{https://iopscience.iop.org/article/10.1088/1361-648X/ae0b1a}{J. Phys.: Condens. Matter \textbf{37}, 405702 (2025).}







\bibitem{Yuan2018}
X.~Yuan, \textit{et al.}, ``Chiral Landau levels in Weyl semimetal NbAs with multiple topological carriers,
''
\href{https://www.nature.com/articles/s41467-018-04080-4}{Nat. Commun. \textbf{9}, 1854 (2018).}


\bibitem{Zhao2021}
P.-L.~Zhao, X.-B.~Qiang, H.-Z.~Lu, and X.~C.~Xie,
``Coulomb Instabilities of a Three-Dimensional Higher-Order Topological Insulator'',
\href{https://journals.aps.org/prl/abstract/10.1103/PhysRevLett.127.176601}{Phys.\ Rev.\ Lett.\ \textbf{127}, 176601 (2021).}



\bibitem{flensberg}
H. Bruus and K. Flensberg, Many-Body Quantum Theory in Condensed Matter Physics: An Introduction,
Oxford Graduate Texts, Oxford University Press, Oxford, 2004. ISBN: 978-0-19-856633-5.

\bibitem{rammer}
J.~Rammer,
\textit{Quantum Field Theory of Non-equilibrium States}
(Cambridge University Press, 2007).

\bibitem{kubo_greenwood}
D.~A.~Greenwood,
``The Boltzmann equation in the theory of electrical conduction in metals,''
\href{https://iopscience.iop.org/article/10.1088/0370-1328/71/4/306}{\textit{Proc. Phys. Soc.} \textbf{71}, 585 (1958).}


\bibitem{BurkovBalents2011}
A.~A.~Burkov and L.~Balents,
``Weyl semimetal in a topological insulator multilayer,''
\href{https://journals.aps.org/prl/abstract/10.1103/PhysRevLett.107.127205}{\textit{Phys.\ Rev.\ Lett.} \textbf{107}, 127205 (2011).}




\bibitem{liang2017multistep}
I. Crassee \textit{et al.},
``{BiTeCl and BiTeBr: A comparative high-pressure optical study}'',
\href{https://doi.org/10.1103/PhysRevB.95.045201}{Phys.\ Rev.\ B {\bf 95}, 045201 (2017)}.

\bibitem{sharma2017multiweyltransport}
G.~Sharma, P.~Goswami, and S.~Tewari,
``{Chiral anomaly and longitudinal magnetotransport in type-II Weyl semimetals}'',
\href{https://doi.org/10.1103/PhysRevB.96.045112}{Phys.\ Rev.\ B {\bf 96}, 045112 (2017)}.






		
	\end{thebibliography}
\end{document}